\documentclass{moriond}

\def\Journal#1#2#3#4{{#1} {\bf #2}, #3 (#4)}

\def\NPB{{\em Nucl. Phys.} B}

\def\PRD{{\em Phys. Rev.} D}

\def\be{\begin{equation}}
\def\ee{\end{equation}}
\def\bea{\begin{eqnarray}}
\def\eea{\end{eqnarray}}

\def\vub{V_{ub}}
\def\vcb{V_{cb}}
\def\btodlnu{B \to D\ell\nu}
\def\btodstlnu{B \to D^*\ell\nu}
\def\costby{\cos\theta_{BY}}
\def\eecl{E_{\rm ECL}^{\rm extra}}
\def\mmiss{M_{\rm miss}^2}
\def\zdiff{z_{\rm diff}}
\def\rd{\mathcal{R}\left(D\right)}
\def\rdp{\mathcal{R}\left(D^+\right)}
\def\rdst{\mathcal{R}\left(D^*\right)}
\def\rdstp{\mathcal{R}\left(D^{*+}\right)}
\def\rddst{\mathcal{R}\left(D^{(*)}\right)}

\newcommand{\Photo}{\includegraphics[height=35mm]{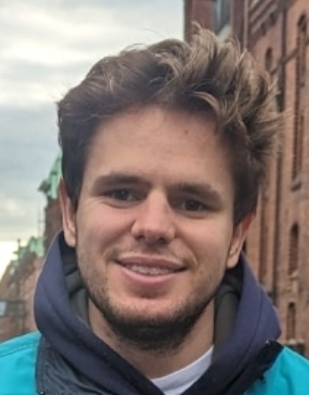}}

\begin{document}
\vspace*{4cm}
\title{Measurements of semileptonic and leptonic beauty meson decays at Belle~II}

\author{Tommy Martinov, on behalf of the Belle~II collaboration}

\address{DESY, Notkestr. 85,\\
Hamburg, Germany}

\maketitle\abstracts{Three recent measurements of leptonic and semileptonic decays of $B$ mesons are presented here. All results use the full Belle~II Run 1 $e^+e^- \to \Upsilon(4S) \to B\bar{B}$ dataset of 365 fb$^{-1}$. We present one leptonic $B$ decay result: a measurement of the $B^+ \to \tau \nu$ branching fraction with a hadronic tagging method; and two semileptonic $B$ decay results: a determination of $|\vcb|$ using $B \to D \ell \nu$ decays with an inclusive tagging method and a test of lepton flavour universality with measurements of $\rdp$ and $\rdstp$ using semileptonic $B$ tagging.} 

\section{CKM matrix elements $|\vcb|$ and $|\vub|$}
The Cabibbo–Kobayashi–Maskawa (CKM) matrix is a $3\times3$ unitary matrix which is the source of all flavour-violating interactions in the quark sector of the Standard Model (SM). The CKM matrix was introduced in order to explain the experimentally observed \textit{charge-parity} violation effects. Precise measurements of the magnitudes of all CKM matrix elements are crucial to test the SM. The two elements known with the lowest precision are $|\vub|$ and $|\vcb|$. The magnitudes of these two elements are usually measured via semileptonic $B$ meson decays either using an exclusive method, where a specific meson is reconstructed (for instance, a $\pi$ to measure $|\vub|$), or an inclusive method where no explicit requirements are applied to the hadronic system. The two methods yield values which differ by about $3\sigma$ for both $|\vub|$ and $|\vcb|$~\cite{hflav}. The value of $|\vub|$ can also be measured via purely leptonic $B$ decays which which have a negligible theoretical uncertainty.

\subsection{Measurement of $|\vcb|$ via $\btodlnu$ decays}
\begin{figure}[h!]
    \centering
    \includegraphics[scale=0.9]{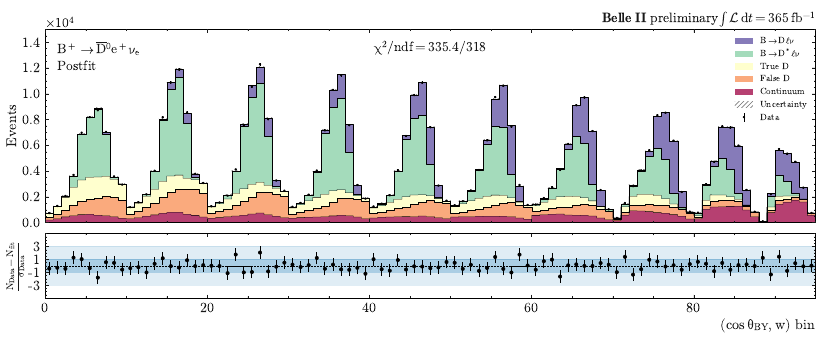}
    \caption{Fitted distribution of $\costby$ in bins of $w$ in the $D^0e^-$ channel. The black points show the Belle~II data. The stacked histograms are simulated events. The true and fake $D$ components respectively represent events with a correctly reconstructed $D$ meson which doesn't stem from a $B \to D^{(*)} \ell \nu$ decay and events with a misreconstructed $D$ meson. The panel at the bottom shows the data -- simulation difference normalised to the total estimated uncertainty.}
    \label{fig:2d_fit}
\end{figure}

\begin{figure}[h!]
    \centering
    \includegraphics[scale=0.8]{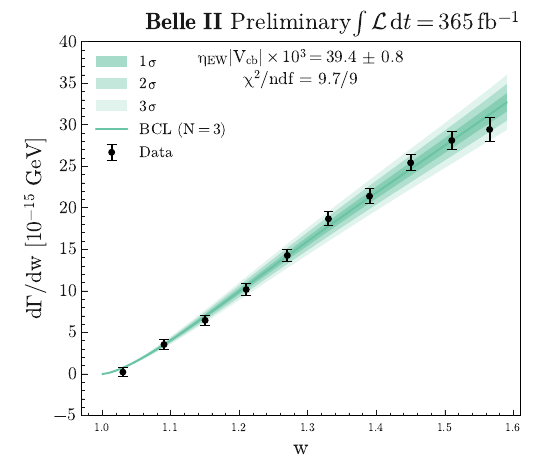}
    \caption{Differential decay rate fit. The black points are the 10 values of $\Delta\Gamma/dw$ extracted from the 2D fit described in the text. The green line represents the central value of the fitted distribution using the BCL form factor parametrisation and the green bands represent the $1\sigma$, $2\sigma$ and $3\sigma$ uncertainties from darker to lighter.}
    \label{fig:bcl_fit}
\end{figure}

The value of $|\vcb|$ is commonly extracted via $\btodstlnu$ transitions which have a larger branching fraction and therefore offer a larger data set to perform the measurement. Nevertheless, studying $\btodlnu$ decays yields several advantages. The $D^*$ meson is reconstructed via its decays to a $D$ meson and a low-momentum $\pi$ whose reconstruction represents the leading systematic uncertainty in $|\vcb|$ measurements, an issue which is avoided when using $\btodlnu$ decays. Furthermore, lattice QCD predictions of $\btodlnu$ form factor parameters are more precise than for $\btodstlnu$ decays. $\btodlnu$ decays can be studied through various parameters such as $q^2$, the invariant mass squared of the leptonic system, defined as $q^2 = (p_\ell + p_\nu)^2$, where $p_\ell$ and $p_\nu$ are the lepton and neutrino four-momenta respectively. Using the $B$ and $D$ meson masses $m_B$ and $m_D$, the recoil parameter $w$ is defined as
\begin{equation}
    w = \frac{m_B^2 + m_D^2 - q^2}{2m_Bm_D}.
\end{equation}
The $\btodlnu$ differential decay rate as a function of $w$ is expressed as~\cite{hqs}
\begin{equation}
    \frac{d\Gamma(\btodlnu)}{dw} = \frac{G_F^2m_D^3}{48\pi^3} (m_B+m_D)^2(w^2-1)^{3/2}\eta_{\rm EW}^2\mathcal{G}^2(w)|\vcb|^2,
\end{equation}
where, $\eta_{\rm EW} = 1.0066 \pm 0.0002$ is the electroweak correction~\cite{etaew} and $\mathcal{G}(w)$ is the $\btodlnu$ form factor. The Belle~II measurement is performed by reconstructing both $B^0$ and $B^+$ decays without explicitly reconstructing the partner $B$ meson stemming from the $e^+e^- \to \Upsilon(4S) \to B\bar{B}$ process. This method, known as \textit{inclusive tagging}, allows to maximise the number of events available for the measurement. Charged and neutral $D$ mesons are reconstructed via the decays $D^+ \to K^- \pi^+ \pi^+$ and $D^0 \to K^- \pi^+$. The signal is extracted with a two-dimensional fit to $w$ and the $\costby$ variable defined as
\begin{equation}\label{equ:costby}
    \costby = \frac{2E^*_BE^*_Y-m_B^2-m_Y^2}{2|p_B^*||p_Y^*|},
\end{equation}
where $Y$ represents the $D\ell$ two-body system and $E^*$ and $p^*$ are the energies and momenta of the $B$ meson and $Y$ system in the centre-of-mass frame. The postfit 2D $w:\costby$ distribution is shown in Figure~\ref{fig:2d_fit}. The fit is simultaneously performed on four separate channels: $D^0e^-$, $D^0\mu^-$, $D^+e^-$ and $D^+\mu^-$ to extract the individual $B^0 \to D^+\ell^-\nu$ and $B^- \to D^0\ell^-\nu$ branching fractions and a lepton flavour universality (LFU) test. The branching fractions are compared to the latest Heavy Flavour Averaging Group (HFLAV) averages~\cite{hflav} in Table~\ref{tab:br} and the LFU test reads: $\mathcal{B}(B \to D e \nu) / \mathcal{B}(B \to D \mu \nu) = 1.02 \pm 0.03$.
\begin{table}[h!]
    \centering
    \caption[B to D l v BF]{$\btodlnu$ branching fractions as extracted from the fit compared to the latest HFLAV averages~\cite{hflav}.}
    \begin{tabular}{|c c c|}
         \hline
         & Belle~II & HFLAV \\
         \hline
        $\mathcal{B}(B \to D^- \ell^+ \nu)$ (\%) & $2.06 \pm 0.12$ & $2.12 \pm 0.06$ \\
        $\mathcal{B}(B \to \bar{D}^0 \ell^+ \nu)$ (\%) & $2.31 \pm 0.10$ & $2.21 \pm 0.06$ \\
         \hline
    \end{tabular}
    \label{tab:br}
\end{table}
The differential decay rate $\Delta\Gamma/dw$ in 10 bins of $w$ is obtained from the same fit. The values $\Delta\Gamma/dw$ are fitted to the differential decay rate expressed using the Bourrely, Caprini, Lellouch (BCL)~\cite{bcl} form factor parametrisation via a $\chi^2$ fit. In addition, lattice QCD constraints computed by the FMILC~\cite{fmilc} and HPQCD~\cite{hpqcd} collaborations are included in the fit. The fitted differential decay rate is illustrated in Figure~\ref{fig:bcl_fit}. From this fit, a value of $|\vcb|$ is extracted,
\begin{equation}
    |\vcb|_{\rm BCL} = (39.2 \pm 0.8) \times 10^{-3}.
\end{equation}
which can be compared to the exclusive and inclusive averages~\cite{hflav}, $|\vcb|_{\rm excl.} = (39.77 \pm 0.46) \times 10^{-3}$ and $|\vcb|_{\rm incl.} = (41.97 \pm 0.48) \times 10^{-3}$.

\subsection{Measurement of the $B^+ \to \tau \nu$ branching fraction}
\begin{figure}[h!]
    \centering
    \includegraphics[scale=0.8]{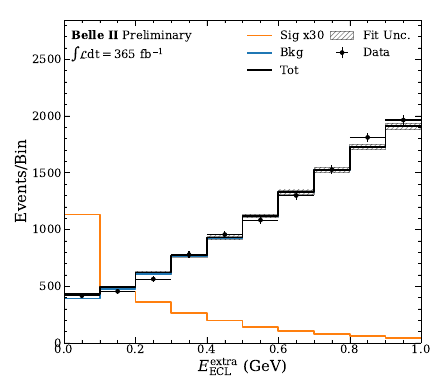}
    \includegraphics[scale=0.8]{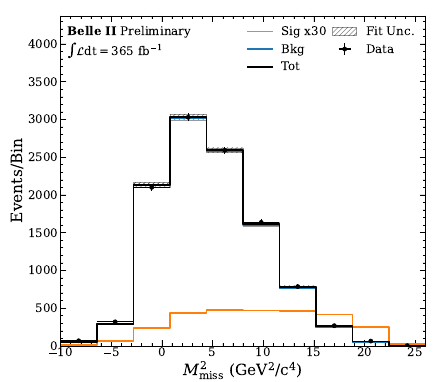}
    \caption[Postfit distributions]{Distributions of the postfit $\eecl$ (left) and $\mmiss$ (right) variables~\cite{gio}. The signal (orange line) is shown multiplied by 30.}
    \label{fig:gio_distributions}
\end{figure}

\begin{figure}[h!]
    \centering
    \includegraphics[scale=0.4]{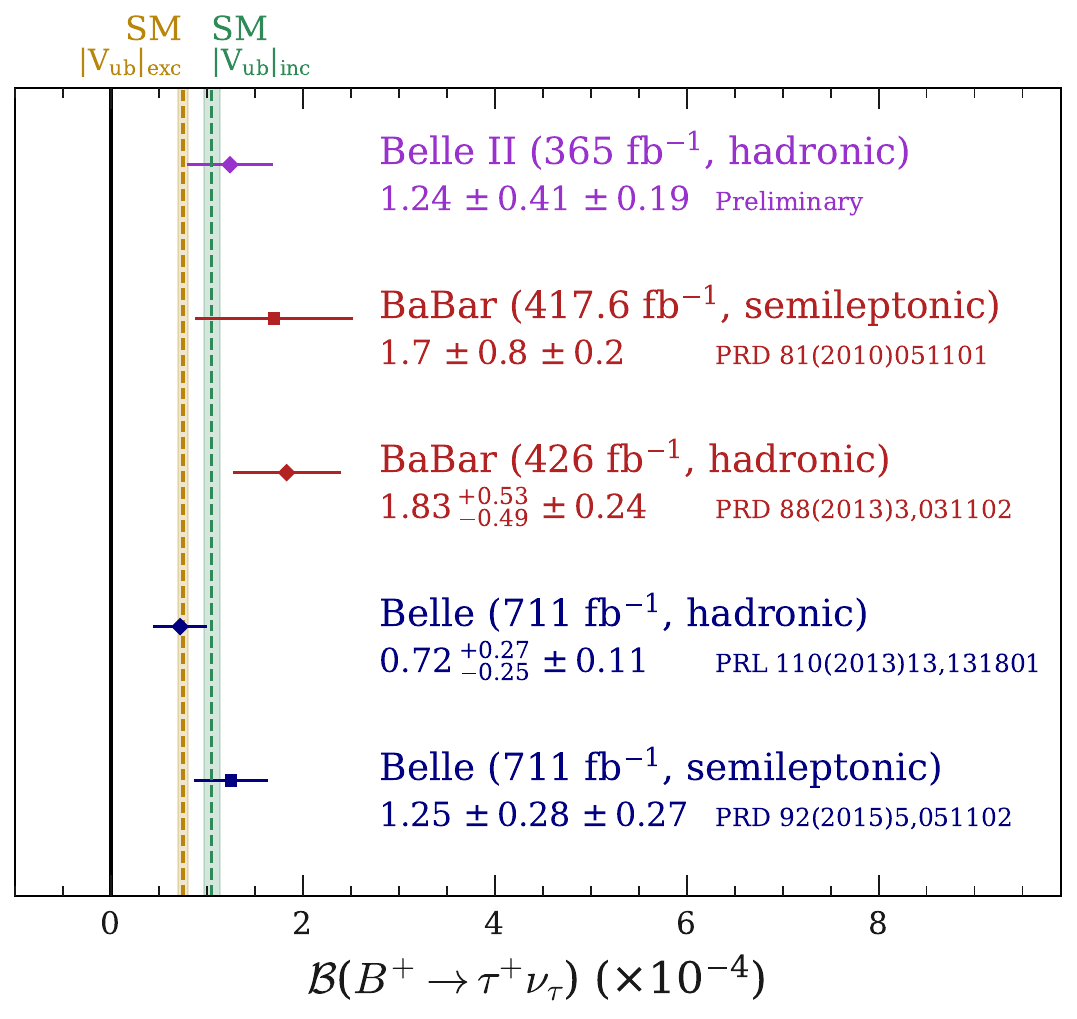}
    \caption[comp]{Comparison between the $B \to \tau\nu$ branching fraction measured by Belle~II (pink), BaBar (red) and Belle (blue)~\cite{gio}. The branching fractions obtained assuming the exclusive (inclusive) value of $|\vub| = (3.42 \pm 0.12)\times 10^{-3}$ ($|\vub| = (4.06 \pm 0.16)\times 10^{-3}$) is shown as a yellow (green) band.}
    \label{fig:Final_Unblinding_}
\end{figure}

Purely leptonic $B$ meson decays represent the theoretically cleanest channel to measure $|\vub|$ but they are helicity suppressed which makes them extremely rare and therefore hard to study with the available datasets. The $B^+ \to \ell^+ \nu$ branching fraction is expressed as
\begin{equation}
    \mathcal{B}(B^+ \to \ell^+\nu_\ell) = \frac{G_F^2m_B}{8\pi}m_\ell^2\left( 1 - \frac{m_\ell^2}{m_B^2} \right)^2 f_B^2|V_{ub}|^2\tau_B,
\end{equation}
where $m_\ell$ is the mass of the lepton, $f_B = 190 \pm 1.3$ MeV~\cite{flag} is the $B$ meson decay constant which can be extracted from lattice QCD and $\tau_B$ is the $B$ meson lifetime. So far, no individual measurement of $\mathcal{B}(B^+ \to \ell^+\nu_\ell)$ has ever crossed the $5\sigma$ discovery threshold. The Belle~II measurement~\cite{gio} is performed by reconstructing the partner $B$ meson in its hadronic decay channels (a method called \textit{hadronic tagging}~\cite{fei}). While it reduces the reconstruction efficiency significantly this strategy is necessary to constrain the event kinematics despite the presence of multiple final-state undetected neutrinos on the signal side. The $\tau$ lepton is reconstructed in four channels to maximise the total efficiency: $\tau \to e\nu\nu$, $\tau \to \mu\nu\nu$, $\tau \to \pi\nu$ and $\tau \to \rho\nu$. The four channels amount to a total of $71.5\%$ of the total $\tau$ decay rate~\cite{pdg}. The signal is extracted from a 2D fit to the $\eecl$ and $\mmiss$ variables which are respectively defined as the total energy from neutral clusters not associated with either $B$ mesons and the squared missing mass of the event. The $\eecl$ distribution is calibrated using three separate control samples to correct $B\bar{B}$ backgrounds, signal decays with $\tau$ leptonic modes and signal decays with $\tau$ hadronic modes. The signal extraction fit is a simultaneous binned maximum likelihood fit of all four signal $\tau$ decays channels. The postfit $\eecl$ and $\mmiss$ distributions are shown in Figure~\ref{fig:gio_distributions}.
The measured branching fraction reads
\begin{equation}
    \mathcal{B}(B^+ \to \tau^+\nu) = (1.24 \pm 0.41 \pm 0.19)\times 10^{-4},
\end{equation}
where the first uncertainty is statistical and the second systematic, yielding a significance of $3.0\sigma$ ($2.7\sigma$ expected significance). The measured branching fraction is compared to previous measurements  in Figure~\ref{fig:Final_Unblinding_}.

\section{Lepton flavour universality tests}
In the SM, the $W$ boson is expected to couple to the three lepton families with the same strength. A violation of LFU would therefore be a clear signature of non-SM couplings with leptons. A key LFU test involves measuring the ratios of branching fractions $\rddst = \mathcal{B}(B \to D^{(*)}\tau\nu)/\mathcal{B}(B \to D^{(*)}\ell\nu)$ with $\ell=e,\mu$. The SM predictions read
\bea
    \rdp &= 0.296 \pm 0.004, \\
    \rdstp &= 0.254 \pm 0.005.
\eea
These ratios have been measured several times by BaBar, Belle, LHCb and now Belle~II. The average of the combined $\rd-\rdst$ value is in tension with the SM expectation at a level of about $3.1\sigma$ (see Figure 26 in the latest HFLAV report~\cite{hflav} for a summary).

\subsection{Test of lepton flavour universality with measurements of $\rdp$ and $\rdstp$}
\begin{figure}[h!]
    \centering
    \includegraphics[scale=0.8]{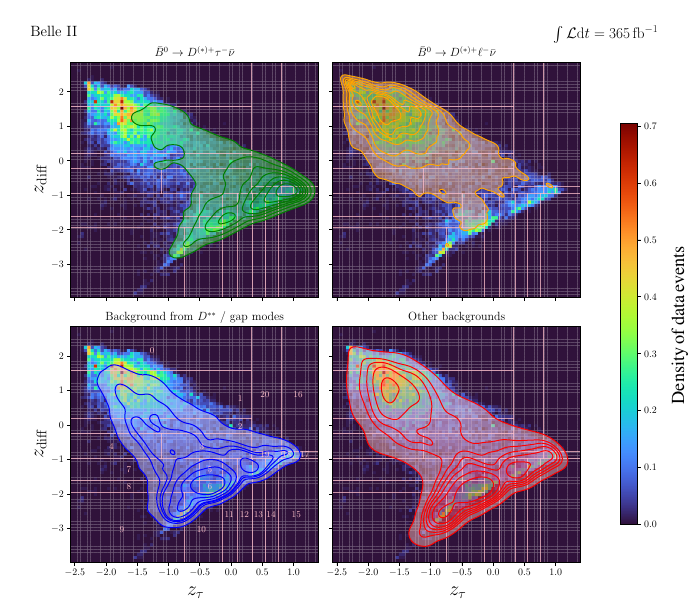}
    \caption[densities]{Densities of the fit categories in the $z_\tau-\zdiff$ plane~\cite{alina}. Semitauonic events are shown in green (top left), semileptonic events in orange (top right), background events from $B \to D^{**}\ell\nu$ decays in blue (bottom left) and other backgrounds in red (bottom right). The density of data events is shown in the background of each figure as a binned histogram. The bin edges chosen for the fit are overlaid on each figure.}
    \label{fig:2d_rd}
\end{figure}

\begin{figure}[h!]
    \centering
    \includegraphics[scale=0.9]{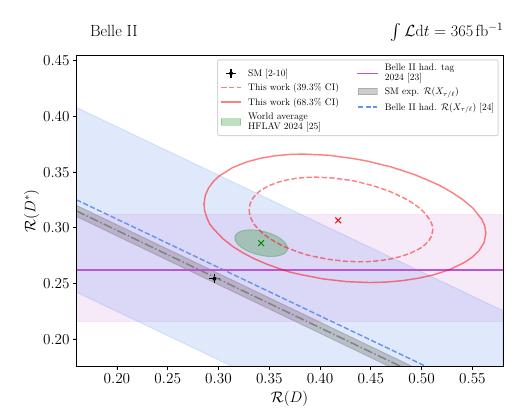}
    \caption[rd]{The measured value of $\rd-\rdst$ (red cross) with the associated $68\%$ (red solid contour) and $39\%$ (red dashed contour) confidence intervals are compared to the SM expectation (black cross) and the world average from the latest HFLAV report~\cite{hflav} (green ellipse)~\cite{alina}. In addition, the Belle~II hadronic tag $\rdst$~\cite{kojima} (pink band) and $\mathcal{R}(X_{\tau/\ell})$~\cite{henrik} (blue band) measurements and the SM $\mathcal{R}(X_{\tau/\ell})$ expectation are shown.}
    \label{fig:rd_comp}
\end{figure}

The first Belle~II combined $\rd-\rdst$ measurement is presented here~\cite{alina}. The measurement is performed with $B^0$ decays by reconstructing the partner $B$ meson in its semileptonic decay channels. This method allows to constrain the kinematics of signal-side decays similarly to the hadronic tagging strategy mentioned earlier. Semileptonic $B$ decays having a larger branching fraction than hadronic decays, the semileptonic tagging method allows to significantly increase the reconstruction efficiency at the cost of less precise kinematic constraints. In this measurement, the $\tau$ is reconstructed via its leptonic decays and $D$ mesons are reconstructed through various decays to charged and neutral $K$ and $\pi$ with a total of 26 decay modes for the tag-side $B$ and 13 for the signal-side $B$. Then, a boosted decision trees (BDT) algorithm is used to classify events in three categories: semitauonic signal events, (light) semileptonic signal events and background events. The most discriminating feature of this BDT is the $\costby$ variable defined in Equation~\ref{equ:costby} with $Y=D\ell$ or $D^*\ell$. Each event is therefore assigned a BDT score $z_\tau$, $z_\ell$ and $z_{\rm bgk}$ and the signal is extracted from a two-dimensional fit to $z_\tau$ and $\zdiff = z_\ell - z_{\rm bkg}$. Events used in the fit are split in four categories: semitauonic, semileptonic, background from $B \to D^{**}\ell\nu$ decays and other backgrounds. The densities of events in these categories in the $z_\tau-\zdiff$ plane are shown in Figure~\ref{fig:2d_rd}. The fit is simultaneously performed on four separate channels: $D^+e^-$, $D^+\mu^-$, $D^{*+}e^-$ and  $D^{*+}\mu^-$.
The values of $\rdp$ and $\rdstp$ extracted from the fit are
\bea
    \rdp &= 0.418 \pm 0.074 \pm 0.051, \\
    \rdstp &= 0.306 \pm 0.034 \pm 0.018,
\eea
with a correlation of $\rho = -0.24$ and where the first uncertainty is statistical and the second systematic. The two values are consistent with the SM expectations within $1.2\sigma$ and $1.6\sigma$ respectively. The combined value is compared to the world average and the SM expectation in Figure~\ref{fig:rd_comp}.

\section{Summary}
Three recent Belle~II results were presented here: a measurement of $|\vcb|$ via $\btodlnu$ decays competitive with previous measurements of $|\vcb|$ via $\btodstlnu$ decays which are usually preferred because of a branching fraction about twice larger; a $B^+ \to \tau\nu$ branching fraction measurement which is competitive with previous measurements and paves the way for future $|\vub|$ measurements with negligible theoretical uncertainty; a combined $\rdp$ and $\rdstp$ measurement which is the first Belle~II result using the semileptonic tagging method and the first Belle~II combined $\rd-\rdst$ measurement.



\section*{References}

\begin{thebibliography}{99}

\bibitem{hflav}S. Banerjee {\it et al}, arXiv:2411.18639
\bibitem{hqs}M. Neubert, \Journal{Phys. Rept}{245}{259}{1994}
\bibitem{etaew}A. Sirlin, \Journal{\NPB}{196}{83}{1982}
\bibitem{bcl}C. Bourrely, I. Caprini and L. Lellouch, \Journal{\PRD}{82}{099902}{2010}
\bibitem{fmilc}J. A. Bailey {\it et al}, \Journal{\PRD}{92}{034506}{2015}
\bibitem{hpqcd}H. Na {\it et al}, \Journal{\PRD}{92}{054510}{2015}
\bibitem{flag}Y. Aoki {\it et al}, arXiv:2411.04268
\bibitem{gio}I. Adachi {\it et al}, arXiv:2502.04885
\bibitem{pdg}S. Navas {\it et al}, \Journal{\PRD}{110}{030001}{2024}
\bibitem{kojima}I. Adachi {\it et al}, \Journal{\PRD}{110}{072020}{2024}
\bibitem{henrik}I. Adachi {\it et al}, \Journal{\PRD}{132}{211804}{2024}
\bibitem{fei}T. Keck {\it et al}, \Journal{Comput. Softw. Big Sci.}{3}{6}{2019}
\bibitem{alina}I. Adachi {\it et al}, arXiv:2504.11220


\end{thebibliography}
\end{document}